\documentclass[
reprint,
prl,
]{revtex4-2}

\usepackage{graphicx}
\usepackage{dcolumn}
\usepackage{bm}

\usepackage{amsmath}
\graphicspath{{./}}     

\graphicspath{ {./ } }
\usepackage{float}

\usepackage{dcolumn}
\usepackage{xcolor}

\begin{document}

\title{Confirming Wave Turbulence Predictions in Rotating Turbulence}

\preprint{APS/123-QED}

\author{Omri Shaltiel}
\altaffiliation{Racah Institute of Physics, The Hebrew University, Jerusalem 91904, Israel}

\author{Omri Gat}%
\affiliation {Racah Institute of Physics, The Hebrew University, Jerusalem 91904, Israel}

\author{Eran Sharon}%
\affiliation {Racah Institute of Physics, The Hebrew University, Jerusalem 91904, Israel}

\date{\today}

\begin{abstract}

Though highly impacting our lives, rotating turbulent flows are not well understood. These anisotropic three-dimensional fluctuating flows are governed by different nonlinear processes, each of which can be dominant in a different range of parameters. More than 20 years ago, Galtier used weak wave turbulence theory (WTT) to derive explicit predictions for the energy spectrum of rotating turbulence. The spectrum is an outcome of forward energy transfer by inertial waves, the linear modes of rotating fluid systems. This spectrum has not yet been observed in freely evolving flows. In this work, we show that the predicted WTT field does exist in steady rotating turbulence, alongside with the more energetic quasi two-dimensional turbulent field. By removing the 2D component from the steady state velocity field, we show that the residual three-dimensional field consists of inertial waves and obeys WTT predictions. Our analysis verifies the dependence of the energy spectrum on all four relevant parameters and provides limits within which WTT predictions hold. These results provide a solid basis for new theoretical and experimental works focused on the coexistence of the quasi 2D field and the inertial waves field and on their interactions.
\end{abstract}

\maketitle

\paragraph{\label{sec:intro}Introduction}
Rapid rotation has competing effects on turbulent flows. On the one hand, it excites and supports the propagation of inertial waves, helical three-dimensional (3D) waves that are driven by Coriolis acceleration \cite{davidson2013turbulence, davidson2024dynamicsbook, alexakis2018cascades}. Thus, as a nonlinear wave system, rotating turbulence might be governed by wave interactions, suggesting that WTT should describe the statistics of such flows. In particular, Galtier derived an explicit prediction for the anisotropic energy spectrum of rotating turbulence using WTT \cite{galtier2003weak}.

On the other hand, experiments and simulations indicate that as the rate of rotation increases, the flow field becomes progressively more confined to the plane perpendicular to the axis of rotation \cite{yarom2013experimental, smith1999transfer, buzzicotti2018inverse, sen2012anisotropy, baroud2003scaling, lamriben2011direct, campagne2014direct, shaltiel2024direct}. In this regime, the flow behaves, in many respects, similarly to two-dimensional (2D) non-rotating turbulence, exhibiting long-lived coherent vortexes and an inverse cascade of energy. Under these conditions, the great majority of kinetic energy is contained in the quasi-2D part of the flow.

In view of these observations, it has not been clear whether a wave turbulence cascade exists in rotating flows. Does the presence of energetic quasi-2D turbulence, which is dominated by inverse energy cascade preclude the wave turbulent cascade of inertial waves? Statistics similar to Galtier's spectrum were measured in simulations, only when modified (local) dynamics was used \cite{yokoyama2021energy} or as an inverse cascade \cite{sharma2018energy,sharma2019anisotropic}. In recent experiments \cite{cortet2020quantitative}, and simulations \cite{le2021evidence}, parts of the Galtier spectrum were observed. However, this observation was made in turbulence where the quasi-2D flow was suppressed, and so far the Galtier spectrum has not been measured in presence of the naturally evolving energetic quasi-2D flow.

In this work, we present a set of experiments in which we decompose the rotating turbulent velocity field into its 2D and residual 3D components. Using this decomposition, we show that in steady state rotating turbulence the 3D WTT flow predicted by Galtier \emph{coexists} alongside the previously observed quasi-2D turbulent field. While the latter is dominant on large scales, the WTT flow is dominant on small scales, driving a forward energy cascade. By analyzing the 3D component of the flow, we confirm the full scaling of the Galtier spectrum, i.e., its dependence on all relevant parameters (defined below).

\paragraph{\label{sec:level2}Theoretical background}
\emph{Rotating turbulent flows} are described in the rotating frame by the rotating Navier-Stokes equation. They are characterized by two dimensionless numbers: the Reynolds number ($\mathrm{Re}$) and the Rossby number ($\mathrm{Ro}$). The Reynolds number is a measure of the ratio of inertial to viscous forces in a fluid flow, and is defined as $\mathrm{Re} = UL/\nu$, where $U$ is a characteristic velocity, $L$ is a characteristic length, and $\nu$ is the kinematic viscosity of the fluid. The Rossby number is a measure of the ratio of inertial to Coriolis forces and is defined as $\mathrm{Ro} = U/(2\Omega L)$, where $\Omega$ is the angular velocity of the system. Rotation has significant effects on turbulence even for $\mathrm{Ro}\simeq1$ \cite{morize2006energy,baqui2015phenomenological}. In this work, we consider strongly rotating turbulent flows, in which the Coriolis force is dominant, i.e. $Re\gg1$ and $\mathrm{Ro}\ll1$.

\emph{Inertial waves} propagate in rapidly rotating incompressible fluids. These waves are small-amplitude solutions to the rotating Navier-Stokes equation, arising from the interplay between inertia and the Coriolis acceleration \cite{greenspan1968theory}. We choose the vertical $z$ direction along the axis of rotation, i.e., $\boldsymbol{\Omega}=\Omega \hat{\boldsymbol{z}}$. The amplitude of plane inertial waves is proportional to $\exp [i(\omega t- \boldsymbol{k} \cdot \boldsymbol{r})]$ where $\omega$ and $\boldsymbol{k}$ are the wave frequency and wave vector, respectively, that satisfy the anisotropic dispersion relation
\begin{equation}
    \omega/2\Omega\ = \pm \cos (\theta)
    \label{eq:dispersion_relation}
\end{equation}
where $\theta$ is the angle between the wave vector $\boldsymbol{k}$ and the rotation axis $\hat{\boldsymbol{z}}$ and $\omega/2\Omega$ is the normalized frequency.

\begin{figure*}
    \includegraphics{ 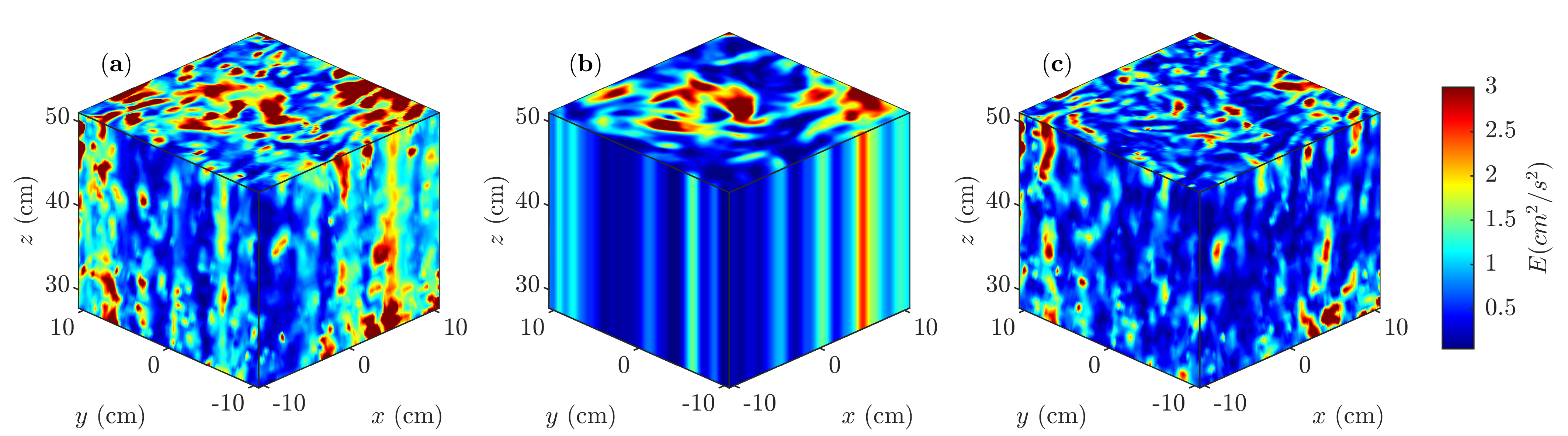}
    \includegraphics{ 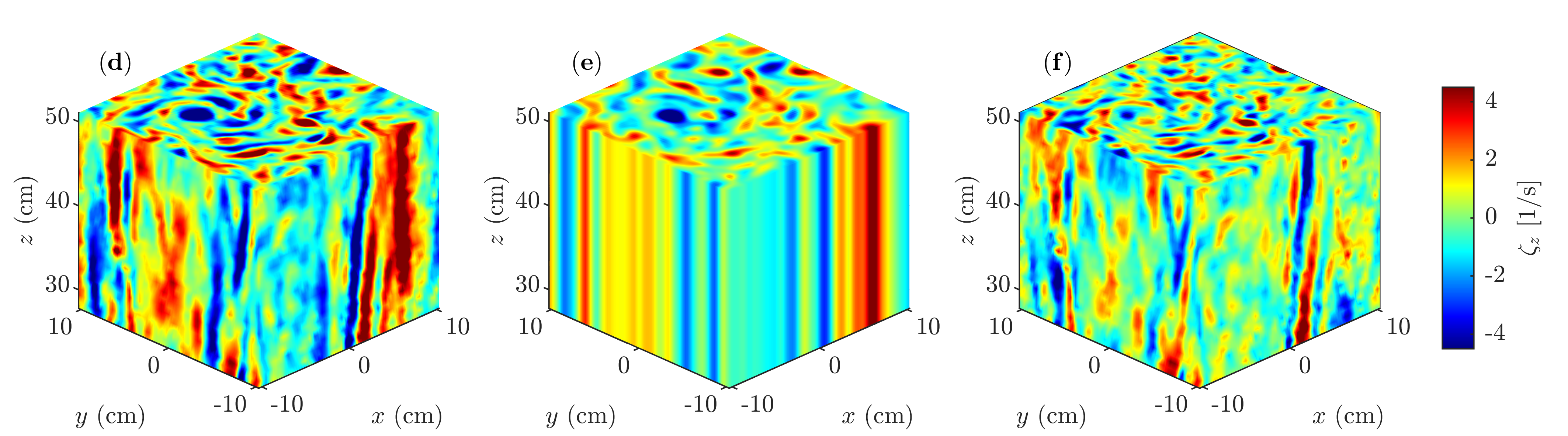}

   \caption{
   A snapshot of the energy density (top row) and $z$ component of vorticity (bottom row) of a rotating turbulent flow field (left), its vertical average $\boldsymbol{v}_{2D}$ (center), and the residual field $\boldsymbol{v}_{3D}$ (right), measured in a cubic domain in an experiment with $\mathrm{Re} \approx 3000$ and $\mathrm{Ro}\approx 0.01$. The full flow field is turbulent and anisotropic, exhibiting structures on a wide range of horizontal scales, and long-range vertical correlations. The 2D field is dominated by disordered, large-scale vortex-like structures that persist for many rotation periods. The 3D field consists of small scales in three dimensions, but with clear anisotropy, showing longer vertical correlation than in the horizontal directions.
   }
    \label{fig:energy_decomposition}
\end{figure*}

\emph{Wave turbulence theory (WTT)} addresses the statistical properties of weakly nonlinear ensembles of waves in large domains \cite{nazarenko2011wave, zakharov1992statistical, galtier2022physicsbook, davidson2024dynamicsbook, newell2011wave}. The theory provides a rigorous analytical framework to derive quantitative predictions for the wave energy spectrum.
WTT has been valuable in understanding energy transfers in systems such as capillary waves \cite{clark2014quantification, falcon2022capilarReview, haudin2016experimentalcapilar} and bending waves in thin elastic plates \cite{mordant2008waveselastic, cobelli2009space, miquel2013transitionwaveselastic, humbert2013waveselastic}.
In the case of rotating turbulence, several theoretical studies have extended WTT to predict the anisotropic energy distribution of inertial waves \cite{nazarenko2011wave, zakharov1992statistical, galtier2022physicsbook, smith1999transfer, galtier2003weak}.

Using WTT formalism, Galtier \cite{galtier2003weak, galtier2023locality} directly calculated that weak interactions of inertial waves can lead to a steady turbulent state. The theory predicts a forward cascade with an anisotropic energy spectrum:

\begin{equation}
   {  \varepsilon_G(k_{r}, k_{z}) \sim \sqrt{\epsilon \Omega} \  k_{r}^{-5/2} k_{z}^{-1/2}}.
    \label{eq:galtier spectrum k_r k_z}
\end{equation}
In this expression, $\epsilon$ denotes the energy injection rate, while $k_z$ and $k_r$ are magnitudes of the vertical and horizontal projections of $k$ respectively. The spectrum \eqref{eq:galtier spectrum k_r k_z} was derived under the assumption that $k_z\ll k_r$. This prediction is expected to apply in the weakly nonlinear regime of rotating turbulence, $Ro\ll1$. At stronger nonlinearity other descriptions may become relevant \cite{nazarenko2011critical}. It is an open question whether the assumptions of WTT can be realized in freely evolving rotating turbulence, i.e., in the presence of the energetic quasi 2D flow, or whether wave turbulence can only be manifested under constrained or controlled conditions \cite{cortet2020quantitative, cortet2020shortcut, le2021evidence, yokoyama2021energy}. The possibility of coexisting quasi-2D and WTT fields is supported by predictions by Scott \cite{scott2014wave} showing that the coupling between inertial waves and strictly 2D modes is weak.  

\paragraph{\label{sec:level2}Experimental System}
We use a rotating plexiglass cylindrical tank of 80 cm diameter and 90 cm height, placed on a rotating table ($\boldsymbol{\Omega} =- \Omega\hat{ \boldsymbol{z}} $, with a maximum rotation rate of $2~Hz$).
The tank is filled with water and covered with a transparent flat lid. Energy is injected at the bottom of the tank by circulating water through an array of outlets and inlets. The energy injection is concentrated at a central wavelength $2\pi/k_{\mathrm{inj}}$, which is a decreasing function of $\Omega$ (see \cite{salhov2019measurements}) down to $\sim5\,$cm at high rotation rates. In the set of measurements shown, $k_{\mathrm{inj}}$ is in the range $0.8~ \textrm{rad}/\text{cm} <k_{\mathrm{inj}}< 1.8 ~\textrm{rad}/\text{cm}$.
Using a vertically scanning horizontal laser sheet, we perform a sweeping particle image velocimetry, and measure the horizontal velocity field, $\boldsymbol{v}_{\perp}(x,y,z,t)$, inside a $\sim 21\times 21\times 24~\text{cm}^3$ volume in the interior of the tank, at a rate (for full volumes) of $21.4$ Hz; the spatial resolution is $0.22 ~$cm horizontally and  $0.7$ cm vertically.
In each experiment, the system is brought to steady state by running it for $\sim 300\text{ s}$ with an angular speed $3.1 \text{ rad}/\text{s} < \Omega < 12.5  \text{ rad}/\text{s}~ i.e. \, (0.5-2~Hz)$ and a constant energy injection rate, corresponding to turbulent flows with Rossby numbers $0.004<\mathrm{Ro}<0.02$ and Reynolds numbers $600<\mathrm{Re}<4500$.

\paragraph{\label{sec:level2}Results}
A snapshot of the kinetic energy $E= (1/2) \,|\boldsymbol{v}(\boldsymbol{r,t})|^2 $ is shown in figure \ref{fig:energy_decomposition}(a). This disordered energy distribution consists of a broad range of scales, both larger and smaller than the injection scale.
We decompose the flow field $\boldsymbol{v}$ into two components: the first is the vertically averaged field, $\bm{v}_{2D}$:
\begin{equation}
    \boldsymbol{v}_{2D}(x,y,t) =  \frac{1}{\Delta h} \int_{-\Delta h/2}^{\Delta h/2} {\boldsymbol{v}} (x,y,z,t)\, dz
\end{equation}
$\Delta h$ is the total measurement height. $\boldsymbol{v}_{2D}$ consists of large, energetic vortical structures (see Fig.~\ref{fig:energy_decomposition}(b), (e)) that meander in the $x,y$ plane (See Supplementary Videos \cite{video_ref,video_ref2}). These large-scale structures are produced by the inverse cascade of energy \cite{shaltiel2024direct, yarom2013experimental, yarom2014experimental, sen2012anisotropy, campagne2014direct, oks2017inverse}.

The second component of the velocity field is the residual $\bm{v}_{3D}$:
\begin{equation}
    \boldsymbol{v}_{3D}(x,y,z,t) =  \boldsymbol{v}(x,y,z,t) - \boldsymbol{v}_{2D}(x,y,t)
\end{equation}

Plotting the energy density and vorticity of this component (Fig.~\ref{fig:energy_decomposition}(c), (f)) reveals that $\boldsymbol{v}_{3D}$ varies on scales much smaller than the scale of variation of $\boldsymbol{v}_{2D}$. Although $\boldsymbol{v}_{3D}$ varies both horizontally and vertically, the snapshots indicate that it is anisotropic, with shorter horizontal and longer vertical scales of variation (See SI for quantitative verification of anisotropy \cite{supp_material}). This anisotropy,  as well as the vertical propagation of energy and vorticity variations, is clearly observable in a videos of the energy and vorticity \cite{video_ref, video_ref2}.

\begin{figure}
    \centering

    \includegraphics[width=0.49\textwidth]{ 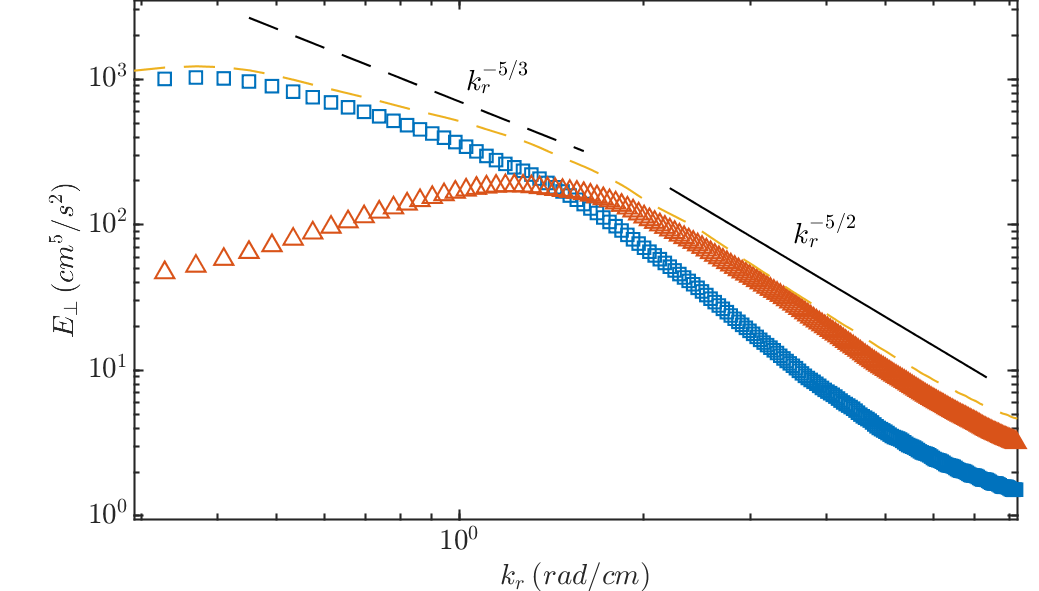}
   \caption{
    Horizontal energy spectra: $E_\perp$ as a function of the horizontally projected wave number $k_r$ of the full velocity field $\boldsymbol{v}$ (yellow dashed line), the vertically averaged flow $\boldsymbol{v}_{2D}$ (blue squares), and the residual field $\boldsymbol{v}_{3D}$ (red triangles). Energy is injected at $k_{r}=k_\textrm{inj}\approx 1.8 \, \textrm{rad}/\textrm{cm}$. For $k_r<k_\textrm{inj}$, the energy density is dominated by $\boldsymbol{v}_{2D}$ which follows a $k_{r}^{-5/3}$ power law. For $k_r>k_\textrm{inj}$ $\boldsymbol{v}_{3D}$ dominates and the spectra follow a $k_{r}^{-5/2}$ power law, consistent with WTT scaling (Eq.\ \eqref{eq:galtier spectrum k_r k_z}).
}
    \label{fig:energy_spatial_spectra}
\end{figure}

The energy spectra $E_{\perp}(k_{r})=\langle  \int |\boldsymbol{v}(\boldsymbol{k}, t)|^2 \,  k_r \, d\phi \rangle_{t,z}$ (see SI \cite{supp_material}) of the full field, the 2D and 3D components, shown in Fig.\ \ref{fig:energy_spatial_spectra}, exhibit two different scaling regimes. At scales larger than the injection scale $k_\textrm{inj}$, the spectrum of the 2D component follows a scaling of $k_{r}^{-5/3}$ \footnote{Due to the limited horizontal measurement range, only part of this scaling is observed. A wider scaling range was measured in 2D measurements at \cite{yarom2013experimental}. }. This regime is generated via the inverse cascade of energy, as previously reported \cite{shaltiel2024direct, yarom2013experimental, yarom2014experimental, sen2012anisotropy, campagne2014direct, oks2017inverse}. At these scales, the vertically averaged $\boldsymbol{v}_{2D}$ field dominates the spectrum, while the residual $\boldsymbol{v}_{3D}$ is negligible.
In contrast, the smaller scales are dominated by $ \boldsymbol{v}_{3D}$, whose energy density scales as $k_{r}^{-5/2}$, consistent with the WTT prediction for a forward cascade [Eq.\ \eqref{eq:galtier spectrum k_r k_z}].

\begin{figure}[b]
    \centering
    \includegraphics[width=0.49\textwidth]{ 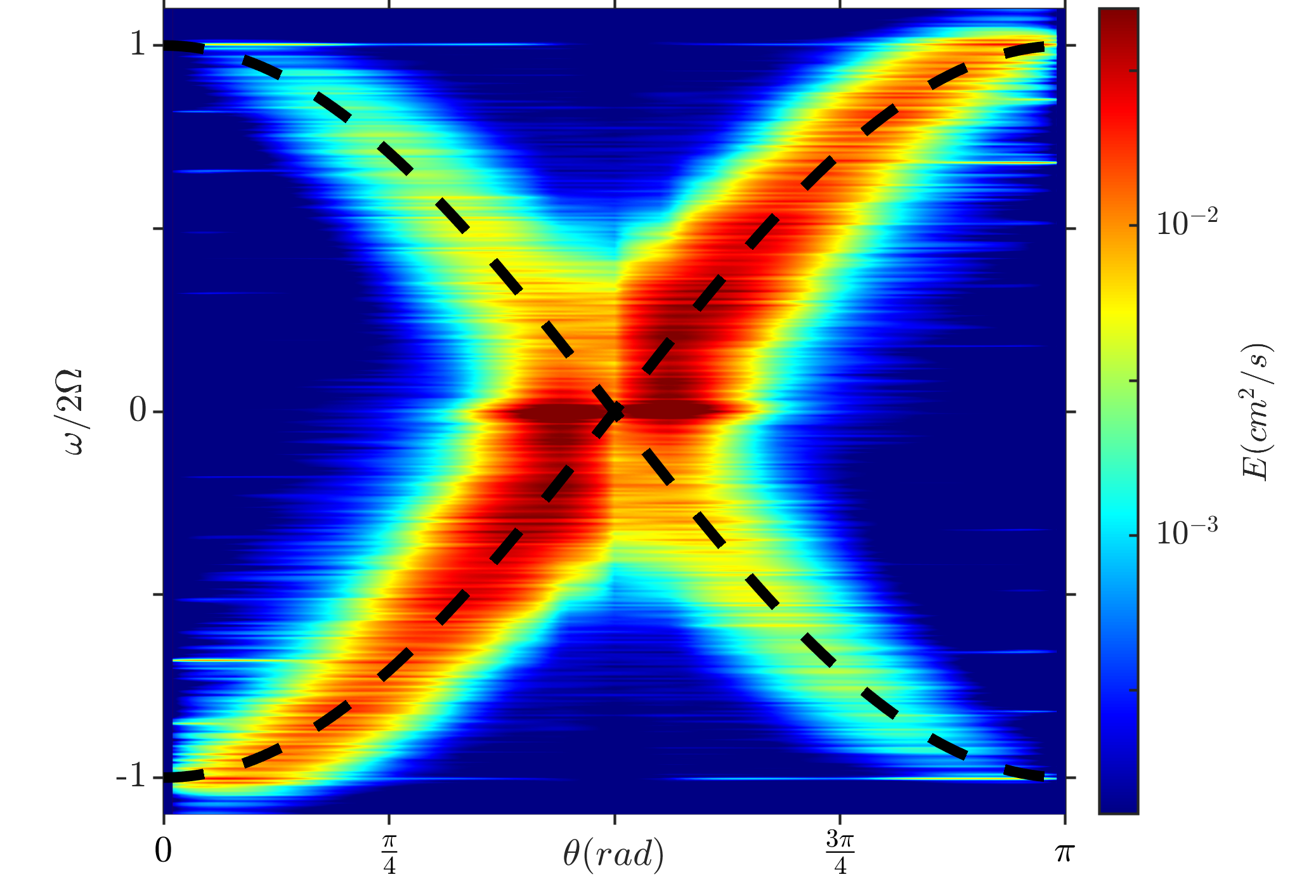}

   \caption{
   The energy density $E(\theta,\omega)$ of the residual flow field $ \boldsymbol{v}_{3D}$ shown as a function of the normalized frequency and angle $\theta$ between the wave vector $\bm{k}$ and the axis of rotation $\bm{\Omega}$. Energy is concentrated along the inertial wave dispersion relation (dashed lines).
    }
    \label{fig:dispersion relation after the decomposition}
\end{figure}

Motivated by these observations, we examine whether the 3D component of the flow consists of inertial waves and whether its energy spectrum is consistent with the full scaling of Eq. (\ref{eq:galtier spectrum k_r k_z}). The 3D velocity field was Fourier transformed in space and time, leading to $\boldsymbol{v}_{3D}(\boldsymbol{k}, \omega)$. The corresponding 4D energy spectrum is defined as $E(\boldsymbol{k}, \omega) = (1/2) \,  |\boldsymbol{v}_{3D}(\boldsymbol{k}, \omega)|^2$ and its average over $\phi$ and small scales (relative to the injection scale) is $E(\theta, \omega)$. Plotting $E(\theta, \omega)$ (Fig \ref{fig:dispersion relation after the decomposition}), we find that the kinetic energy is localized along the dispersion relation Eq.\ (\ref{eq:dispersion_relation}). This spectral behavior is a direct confirmation that $\boldsymbol{v}_{3D}$ consists of inertial waves. In contrast to the spectrum of the full velocity field (see \cite{yarom2014experimental, yarom2017experimental, shaltiel2024direct}), the residual 3D flow does not exhibit a pileup of energy near $\theta\rightarrow \pi/2$ (corresponding to very low frequencies), showing that the slow quasi-2D modes are separated from the 3D flow field. 

The scaling of the energy spectrum in Eq.\ \eqref{eq:galtier spectrum k_r k_z} is highly anisotropic, and its validation requires an independent measurement of the energy density as a function of both $k_z$ and $k_r$. Since $\bm{v}_{3D}$ consists of inertial waves (Fig.3), we use the dispersion relation [Eq.~\eqref{eq:dispersion_relation}] to express $k_z$ in terms of $\omega$, for which we have excellent resolution (see SI~\cite{supp_material} for additional details). Rewriting the dispersion relation [Eq.~\eqref{eq:dispersion_relation}] in terms of the normalized frequency gives $\omega/(2\Omega) = k_z/k\approx k_z/k_r$, the mixed wave vector-frequency energy spectrum associated with Eq.~\eqref{eq:galtier spectrum k_r k_z} \footnote{including a Jacobian, which does not appear in Eq.~\eqref{eq:galtier spectrum k_r k_z}. So this is the energy per unit mass per $k_r$ per $\omega$.} becomes

\begin{equation}
    { E_G(k_{r},\omega) \sim \sqrt{\epsilon/\Omega} \, k_{r}^{-2} (\omega/2\Omega)^{-1/2}\ }.
    \label{eq:galtier spectrum k_r omega}
\end{equation}

\begin{figure}[!t]
    \centering
    \includegraphics[width=0.49\textwidth]{ 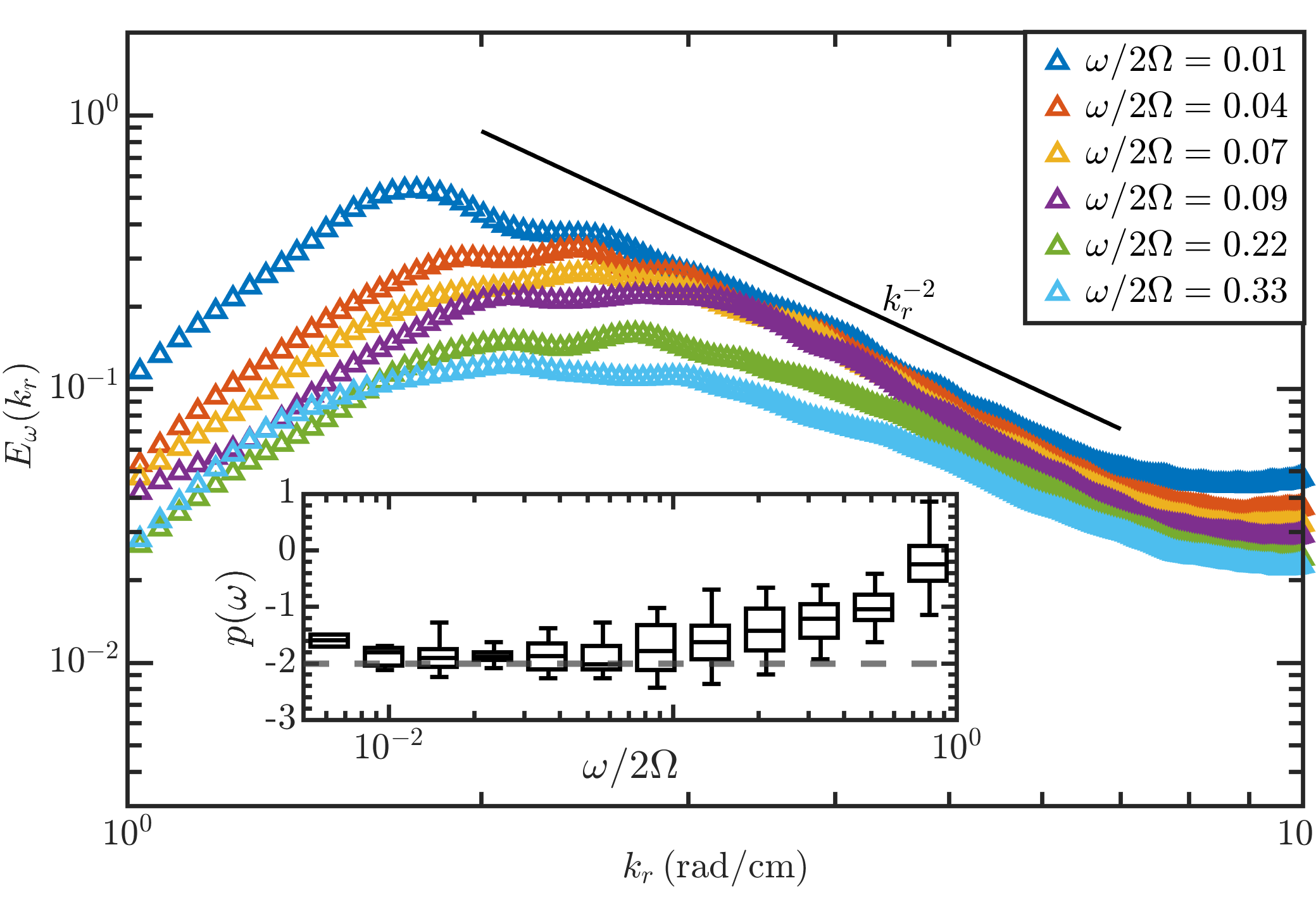 }

   \caption{
    Energy density $E_{\omega}(k_r)$ of the residual 3D flow, shown as a function of the horizontally projected wavenumber $k_r$ for several fixed values of frequency $\omega$. At low frequencies, the spectra are consistent with the $k_r^{-2}$ scaling predicted by WTT, while at higher frequencies clear deviations from this scaling appear. Inset: fitted exponent $p(\omega)$ from power-law fits $E_{\omega}(k_r)\propto k_r^{p}$ to the same spectra, pooled over all datasets and grouped into $12$ logarithmically spaced bins in $\omega/(2\Omega)$. A plateau near $p(\omega)\approx -2$ is observed for $10^{-2}\lesssim \omega/(2\Omega)\lesssim 10^{-1}$.
    }
    \label{fig:energy_of_k_r}
\end{figure}

We analyze the measured two-dimensional spectrum $E(k_r,\omega)$  by examining its behavior at fixed frequency slices \( E_{\omega}(k_r) \equiv E(k_r, \omega=const) \). We plot $E_{\omega}(k_r)$ for a broad frequency range (Fig.\ \ref{fig:energy_of_k_r}). We find a $ k_r^{-2}$ scaling for low frequencies, consistent with Eq. (\ref{eq:galtier spectrum k_r omega}). This scaling is expected to hold only in the limit $k_r\gg k_z$, corresponding to small normalized frequencies $\omega \ll 2\Omega$. Indeed, analyzing six different experiments, we find that for $\omega/(2\Omega)$ of order ($10^{-2}$) to $(10^{-1})$ the best power law fit to the data is $k_r^{-2} $ and the fitting error is small (Fig.\ \ref{fig:energy_of_k_r} inset, and supplemental figure S4-S7 show additional data \cite{supp_material}). For higher frequencies WTT predictions fail.

\begin{figure}[!t]
    \centering
    \includegraphics[width=0.49\textwidth]{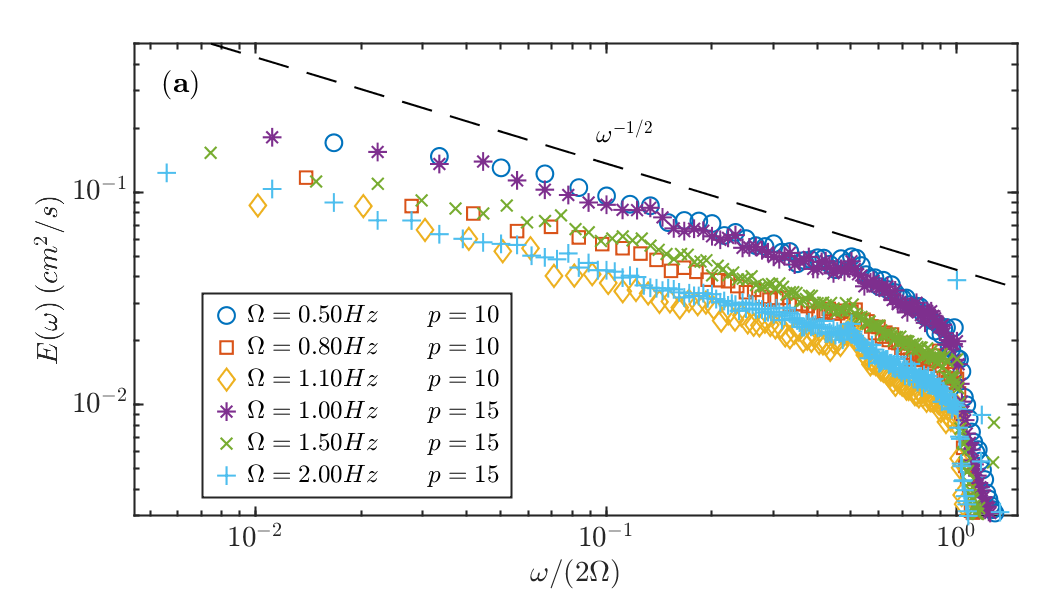}
    \includegraphics[width=0.49\textwidth]{ 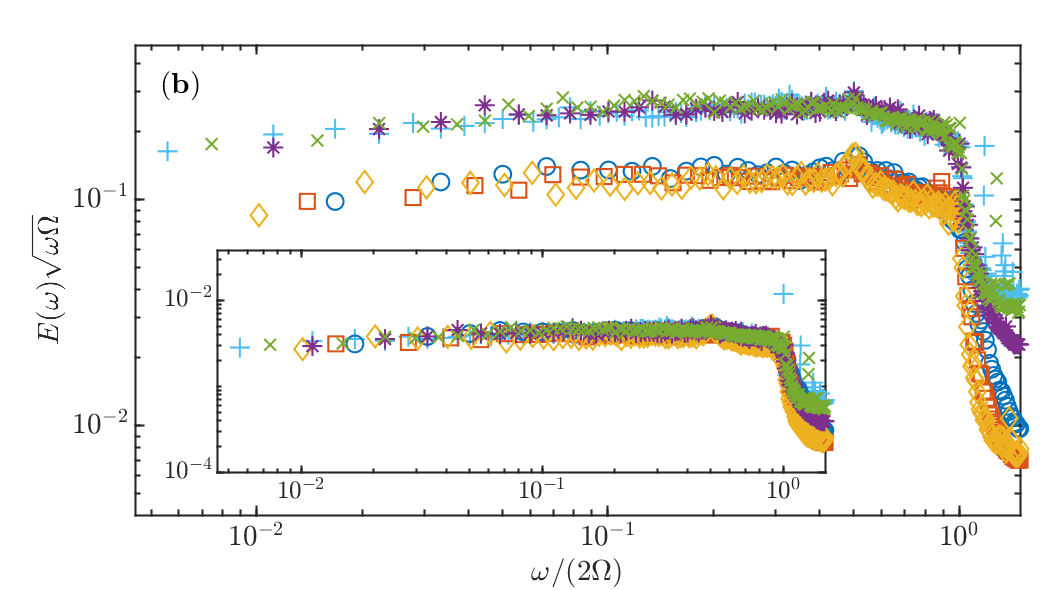}

    \caption{
    (a) Temporal energy spectrum of the residual 3D velocity field $\boldsymbol{v}_{3D}$ as a function of the normalized frequency $\omega/(2\Omega)$ for several experiments with different rotation rates $\Omega$ and injection pressure $p$. For \(\omega/(2\Omega)\) in the intermediate low-frequency range, of order \(10^{-2}\) to \(10^{-1}\), the data are consistent with \(E(\omega)\sim (\omega/2\Omega)^{-1/2}\), as predicted by WTT. (b) The same data shown in panel (a) compensated by $\sqrt{\omega\Omega}$ in the main plot, and additionally by a factor $p^{-3/2}$ in the inset. The good data collapse confirms the scaling of the energy spectrum with frequency, angular velocity, and forward energy flux.
    }
    \label{fig:time_substracted spectrum}
\end{figure}

By integrating $E( \theta , \omega )$ (Fig. \ref{fig:dispersion relation after the decomposition}) over $\theta$, we obtain the temporal energy spectrum $E(\omega)$ of the residual 3D flow. Fig.~\ref{fig:time_substracted spectrum} shows $E(\omega)$ as a function of the normalized frequency $\omega/(2\Omega)$ for experiments with several rotation rates and two different injection pressures $p$, corresponding to different energy injection rates, $\epsilon$. The frequency spectra obey a power law $E(\omega)\  \sim A \,(\omega/ 2\Omega)^{-1/2}$, in agreement with Eq.\ \eqref{eq:galtier spectrum k_r omega}.

We can further test Eq. \eqref{eq:galtier spectrum k_r omega} by examining the dependence of the prefactor $A$ on $\Omega$ and $\epsilon$. Fig.\ \ref{fig:time_substracted spectrum}(b) shows the spectrum $E(\omega)$, compensated by $\sqrt{\omega\Omega}$. After this rescaling, the data collapse onto two separate tightly distributed horizontal "lines". Each of the two lines contains data obtained at different rotation rates, $\Omega$, and they are separated only by the injection pressure $p$. This confirms the $A\sim{\Omega^{-1/2}}$ scaling. Although in our experiments we cannot accurately determine the value of the forward energy flux $\epsilon$, we utilize the fact that it must increase monotonically with the total power injected into the tank, which scales like $p^3$ \footnote{ The kinetic energy injected power is $\dot{m}v^2$, the rate of injected mass times the velocity square. It is therefore proportional to the flow rate in third power, which is proportional to $p^3$ in our system.}. Indeed, additional rescaling of the data by $p^{3/2}$ leads to data collapse around a single horizontal line (inset of Fig.\ \ref{fig:time_substracted spectrum}(b))over the range $10^{-2}\lesssim \omega/(2\Omega)\lesssim 10^{-1}$. This completes the full verification of WTT predictions for the energy spectrum.

\paragraph{Conclusions}
There are two competing paradigms for rapidly rotating weakly turbulent flows. The first rests on the observation that the rotation inhibits fluid flow parallel to its axis, making the flow similar to 2D turbulence with an inverse cascade of energy. The second starts from the observation that rotating flows support the propagation of waves, on which wave turbulence can develop; WTT then predicts an anisotropic forward energy cascade.

Previous experiments and simulations have mostly revealed evidence supporting the 2D turbulence paradigm, together with evidence that inertial waves do exist and play a role in energy transport. The experiments presented here show that the two paradigms are not mutually exclusive. Quasi-geostrophic modes transport energy to larger scales, whereas other modes transport energy to smaller scales. The key enabling factor for this discovery was the decomposition of the velocity field into its 2D and 3D components, allowing for the analysis of the wave field separately from the highly energetic large scale flow.

In this manner we were able to identify for the first time two coexisting distinct scaling regimes in energy spectrum of rotating turbulence, and to verify the prediction of inertial wave turbulence theory, made by Galtier more than 20 years ago. The WTT spectrum is anisotropic, and by relating the wavenumber and frequency spectra we were able to verify both the parallel and perpendicular scaling, as well as the parametric scaling in a wide range of flow parameters with $10^{3}\lesssim\mathrm{Re}\lesssim10^{4}$, and $10^{-3}\lesssim\mathrm{Ro}\lesssim 10^{-2}$. The results demonstrate conclusively that an inertial wave cascade is realized in rotating turbulent flows.

Interestingly, even though we identified the inertial wave cascade in the 3D flow component, this anisotropic cascade is actually carried by modes with wavevectors that are nearly perpendicular to the rotation axis, with $k_z\ll k_r$. This limit, which is one of the theoretical conditions of validity, is consistent with the fact that the observed Galtier scaling is found only for modes with $[\omega/(2\Omega)\sim k_z/k_r \ll 1]$, namely in an intermediate range of order \(10^{-2}\) to \(10^{-1}\).

The results presented in this work suggest that rotating turbulence is a double cascade system. The total injected energy is split into two different coexisting cascades. The interaction between the two components of the flow is weak, consistently with the predictions made by Scott \cite{scott2014wave}. Further theoretical and experimental work is needed in order to determine the interplay between the quasi-2D and the wave components of the flow, as well as the dynamics that govern high frequency modes. 
\bibliographystyle{unsrt}

\bibliography{my_bib_file}
\end{document}